\documentclass[12pt]{article}

\usepackage{graphicx}
\usepackage[margin=1.0in]{geometry}
\usepackage{amssymb}
\usepackage[affil-it]{authblk}

\begin{document}

\title{High-energy neutrino astronomy\\
       and the Baikal-GVD neutrino telescope}


\author[a]{A.D.~Avrorin}
\author[a]{A.V.~Avrorin}
\author[a]{V.M.~Aynutdinov}
\author[b]{Z.~Bard\'{a}\v{c}ov\'{a}}
\author[c]{R.~Bannasch}
\author[d]{I.A.~Belolaptikov}
\author[d]{V.B.~Brudanin}
\author[e]{N.M.~Budnev}
\author[a]{G.V.~Domogatsky}
\author[a]{A.A.~Doroshenko}
\author[e]{A.N.~Dyachok}
\author[a]{Zh.-A.M.~Dzhilkibaev}
\author[d]{V.~Dik}
\author[b,d]{R.~Dvornick\'{y}}
\author[b]{E.~Eckerov\'{a}}
\author[d]{T.V.~Elzhov}
\author[f]{L.~Fajt}
\author[g]{S.V.~Fialkovski}
\author[e]{A.R.~Gafarov}
\author[a]{K.V.~Golubkov}
\author[d]{N.S.~Gorshkov}
\author[e]{T.I.~Gress}
\author[d]{R.A.~Ivanov}
\author[d]{M.S.~Katulin}
\author[c]{K.G.~Kebkal}
\author[c]{O.G.~Kebkal}
\author[d]{E.V.~Khramov}
\author[d]{M.M.~Kolbin}
\author[d]{K.V.~Konischev}
\author[h]{K.A.~Kopa\'{n}ski}
\author[d]{A.V.~Korobchenko}
\author[a]{A.P.~Koshechkin}
\author[i]{V.A.~Kozhin}
\author[a]{M.K.~Kryukov}
\author[d]{M.V.~Kruglov}
\author[g]{V.F.~Kulepov}
\author[a]{M.B.~Milenin}
\author[e]{R.R.~Mirgazov}
\author[d]{D.V.~Naumov}
\author[d]{V.~Nazari}
\author[h]{W.~Noga}
\author[a]{D.P.~Petukhov}
\author[d]{E.N.~Pliskovsky}
\author[j]{M.I.~Rozanov}
\author[d]{V.D.~Rushay}
\author[e]{E.V.~Ryabov}
\author[a]{G.B.~Safronov}
\author[d]{B.A.~Shaybonov}
\author[a]{M.D.~Shelepov}
\author[b,d,f]{F.~\v{S}imkovic}
\author[i]{A.V.~Skurikhin}
\author[d]{A.G.~Solovjev}
\author[d]{M.N.~Sorokovikov}
\author[f]{I.~\v{S}tekl}
\author[d]{E.O.~Sushenok}
\author[a]{O.V.~Suvorova}
\author[e]{V.A.~Tabolenko}
\author[e]{B.A.~Tarashansky}
\author[d]{Y.V.~Yablokova}
\author[c]{S.~Yakovlev}
\author[a]{D.N.~Zaborov\thanks{Speaker.}}

\affil[a]{Institute for Nuclear Research, Russian Academy of Sciences, Moscow, Russia}
\affil[b]{Comenius University, Bratislava, Slovakia}
\affil[c]{EvoLogics Gmbh, Berlin, Germany}
\affil[d]{Joint Institute for Nuclear Research, Dubna, Russia}
\affil[e]{Irkutsk State University, Irkutsk, Russia}
\affil[f]{Czech Technical University in Prague, Prague, Czech Republic}
\affil[g]{Nizhny Novgorod State Technical University, Nizhny Novgorod, Russia}
\affil[h]{Institute of Nuclear Physics of Polish Academy of Sciences (IFJ~PAN), Krak\'{o}w, Poland}
\affil[i]{Moscow State University, Moscow, Russia}
\affil[j]{St.~Petersburg State Marine Technical University, St.Petersburg, Russia}

\date{}

\maketitle

\begin{abstract}
Neutrino astronomy offers a novel view of the non-thermal Universe and is complementary to other astronomical disciplines. The field has seen rapid progress in recent years, including the first detection of astrophysical neutrinos in the TeV-PeV energy range by IceCube and the first identified extragalactic neutrino source (TXS 0506+056). Further discoveries are aimed for with new cubic-kilometer telescopes in the Northern Hemisphere: Baikal-GVD, in Lake Baikal, and KM3NeT-ARCA, in the Mediterranean sea. The construction of Baikal-GVD proceeds as planned; the detector currently includes over 2000 optical modules arranged on 56 strings, providing an effective volume of 0.35 km$^3$. We review the scientific case for Baikal-GVD, the construction plan, and first results from the partially built array.

\end{abstract}

\section{Introduction}
The neutrino provides us with unique opportunities for exploring the Universe.
The neutrino is able to escape from dense environments, it travels unimpeded through gas and dust clouds, and
it does not interact with infrared or CMB radiation on its way to the observer (unlike photons).
As a result, the cosmos is practically transparent to neutrinos.\footnote{We leave aside the beyond-standard-model scenarios such as, e.g., resonant neutrino self-interactions.}
The neutrino is not affected by magnetic fields, and thus it always points to its source.
Finally, it is stable (does not decay).
These properties make the neutrino an excellent astrophysical messenger, complementary to photons and gravitational waves.
In a typical hadronic scenario, high energy neutrinos are produced in the decays of pions and muons.
Hence the neutrino allows one to trace the production and acceleration sites of cosmic rays.

In the GeV-TeV energy range the neutrino sky is mostly dominated by atmospheric neutrinos -- a side-product of cosmic ray interactions in the Earth's atmosphere.
This atmospheric neutrino background limits the sensitivity of Earth-based neutrino telescopes to astrophysical neutrinos.
With the atmospheric neutrino flux falling steeply with energy, the preferred energy band for observing the astrophysical neutrinos appears to be above $\sim$~30~TeV.
At these energies, typical rates of astrophysical neutrino interactions are just a few events per year per cubic kilometer, necessitating the construction of huge ($\gtrsim$ 1 km$^3$) neutrino telescopes.
Such a large detector volume can be practically achieved using a natural site only.
Particularly suited for that purpose are deep-sea, deep-lake or deep-ice sites characterized by high light transparency, allowing for the use of the Cherenkov detection technique at large scales \cite{Markov1960}.
The installations using this technique for high energy neutrino detection are known as neutrino telescopes.
An optimal neutrino sensitivity is generally obtained for upward-going neutrino.
The detection of downward-going neutrino is hampered by the atmospheric muon background.

Currently there are 4 large neutrino telescopes in operation or under construction around the world: the IceCube detector \cite{IceCube} at the South Pole, the ANTARES \cite{ANTARES} and KM3NeT \cite{KM3NeT_LoI} telescopes in the Mediterranean sea, and the Baikal-GVD telescope at Lake Baikal.
These telescopes have provided us with an initial impression of the high-energy neutrino Universe,
which needs to be refined with further observations.

This contribution has two parts: first, we briefly review the current status of high-energy neutrino astronomy; second, we present the scientific case for Baikal-GVD, construction plan, and first results from the partially built experiment.

\section{High energy neutrino astronomy today}
The TeV-PeV neutrino telescopes presently in operation or under construction are IceCube, ANTARES, KM3NeT, and Baikal-GVD.
All these telescopes use the same neutrino detection principle -- detecting Cherenkov light from secondary particles produced in neutrino interactions.
They also share many of the same technical approaches and features.
Each of them can be seen as an array of optical sensors arranged on vertical strings to form a 3D lattice in a transparent medium.
The optical sensors measure the time and intensity of the Cherenkov light pulses.
The times of the pulses are used to reconstruct the neutrino direction, and the amplitudes provide a measure of the neutrino energy.
Typically, events are classified into two event typologies: tracks, which correspond to $\nu_\mu$ charged current (CC) interactions; and showers, which correspond to $\nu_e$ CC, $\nu_\tau$ CC, and all-flavour neutral current (NC) interactions.

The IceCube detector \cite{IceCube} occupies a 1 km$^3$ volume of antarctic ice at the South Pole. 
It consists of 5160 optical sensors arranged on 86 vertical strings.
The array was completed in 2010. 
In addition to the main array, IceCube has a densely-instrumented core for GeV neutrino studies (DeepCore) and a surface array for air shower detection (IceTop).
IceCube is currently the most sensitive neutrino telescope in the world.
However, its location at the South Pole implies a reduced sensitivity for sources in the Southern celestial hemisphere.
The angular resolution of IceCube is limited to some extent by significant scattering of optical light on ice non-uniformities.

The ANTARES detector \cite{ANTARES} is a 0.02 km$^3$ detector installed in the Mediterranean sea near the French coast. ANTARES consists of 885 optical sensors arranged in triplets along 12 strings. ANTARES has been in operation since 2008 and is expected to be decommissioned soon, handing it duties over to KM3NeT. 
Due to its location in the Northern Hemisphere, ANTARES provides a sky coverage complementary to IceCube.
In particular, the Galactic center is well within ANTARES's field of view.

The KM3NeT-ARCA detector \cite{KM3NeT_LoI} is a 1 km$^3$ volume detector under construction in the Mediterranean sea near Sicily (Italy). When completed, it will consist of 4140 optical sensors installed on 230 strings. KM3NeT uses a highly efficient multi-PMT design for its optical sensors.
As of writing this manuscript, only the first 3 strings have been known to operate in ARCA.
The KM3NeT-ORCA detector, under construction near the French Mediterranean coast, is a densely instrumented version of ARCA, optimized for the neutrino mass hierarchy measurement \cite{KM3NeT_LoI}. Due to the relatively small volume ($\sim$ 0.008 km$^3$), ORCA has a limited sensitivity to TeV-PeV neutrinos.
KM3NeT will provide a field of view complementary to that of IceCube.
Furthermore, thanks to excellent optical properties (low absorption and low scattering) of the deep Mediterranean water, KM3NeT will provide a substantial improvement over the angular resolution of IceCube.

The Baikal-GVD detector is a 1 km$^3$ detector under construction in Lake Baikal (Russia).
Baikal-GVD is competing with KM3NeT-ARCA for the role of the largest neutrino telescope in the Northern Hemisphere.
Further details on Baikal-GVD are provided in the next section.

The recent development of neutrino astronomy has been marked by several notable discoveries and observations.
First came the discovery of a diffuse astrophysical neutrino flux by IceCube \cite{IceCube_diffuse}.
The statistical significance of the excess over the atmospheric neutrino expectations has been steadily growing as more IceCube data became available.
In a recent paper by IceCube \cite{IceCube_diffuse_2020} a 10 $\sigma$ excess is reported using a cascade analysis of 4+2 yrs data.
The IceCube data are generally compatible with a uniform all-sky distribution, with some possible deviations due to a few putative point-like sources.
The flavour composition of the diffuse flux ($\nu_e$:$\nu_\mu$:$\nu_\tau$) is so far compatible with the standard prediction of 1:1:1 after oscillations.
The diffuse flux is confirmed, though with a low statistical significance of 1.6 $\sigma$, by ANTARES \cite{ANTARES_diffuse}.
The origin of the diffuse neutrino flux is so far unknown and is subject to much debate.

Another prominent discovery was the observation of a neutrino event in spatial coincidence with the blazar TXT~0506+056 while in flare state observed in the $\gamma$-ray band \cite{IceCube_TXS0506}.
The neutrino event, recorded on September 22, 2017, was a track-like event reconstructed just 0.1$^\circ$ off the position of TXT~0506+056.
The event itself had a moderate statistical significance.
However an analysis of the earlier IceCube data revealed a 3.7 $\sigma$ excess of events from the same source in the 2015 data.
Taken together, these observations provide a strong evidence for high-energy neutrino emission from TXT~0506+056.

Upper limits have been obtained on neutrino emission from various objects and object classes, including active galactic nuclei, gamma-ray bursts, supernova remnants, and other.
The discovery potential of neutrino astronomy is fully unlocked through combining the neutrino data with data obtained using other messengers - photons (of any wavelength), gravitational waves, and, possibly, cosmic rays.
This is known as the multi-messenger astronomy.

\section{Baikal-GVD status and first results}
The Baikal Gigaton Volume Detector (Baikal-GVD) is a cubic-kilometer scale underwater neutrino detector currently under construction in Lake Baikal, Russia.
The experiment is aimed to provide observations of the neutrino sky with a sensitivity similar to that of IceCube, covering the Southern sky which is poorly visible for IceCube.
Simultaneous operation with ANTARES, KM3NeT and IceCube allows for continuous monitoring of transient phenomena over the full sky and improved all-sky combined sensitivity.

The Baikal-GVD detector site is located in the southern basin of Lake Baikal at 51$^\circ$\,46'\,N 104$^\circ$\,24'\,E, 3.6 km offshore and 1366~m deep.
The light absorption length in the deep lake water reaches 22~m.
The light scattering is between 30~m and 50~m, i.e. relatively small compared to absorption.
The lake is covered with thick ice (up to $\approx$ 1~m) from February to mid-April,
providing a convenient solid platform for detector deployment and maintenance operations.

\begin{figure}
  \centering
  \includegraphics[height=8.5cm]{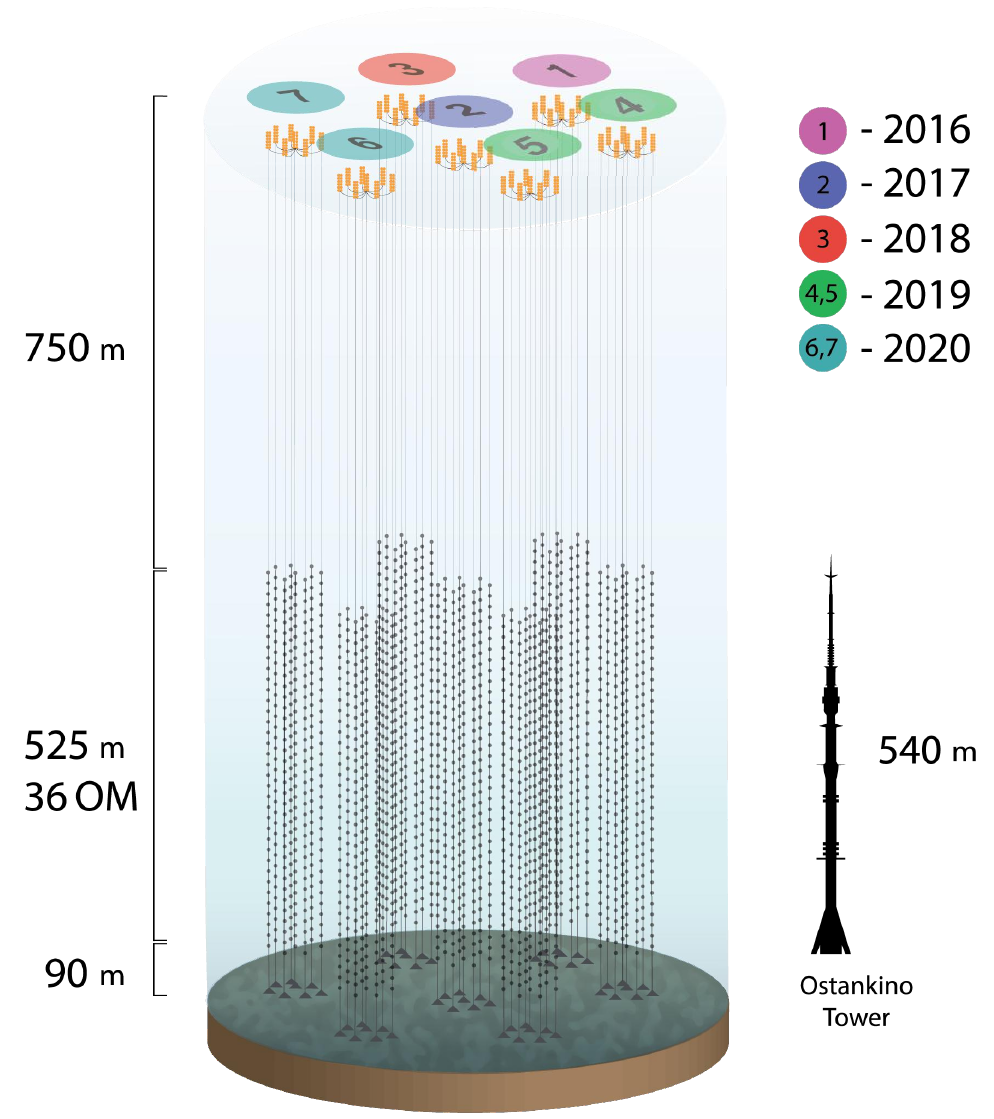}
  \includegraphics[height=8.5cm]{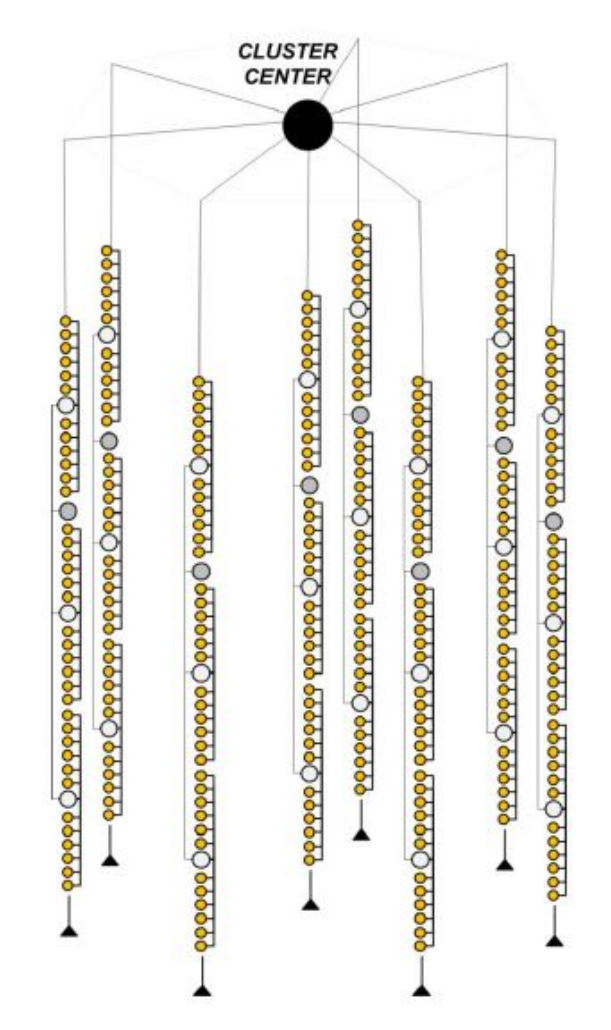}
  \caption{Left: Schematic view of the Baikal-GVD detector (see text).
           Right: The Baikal-GVD cluster layout (vertical proportions compressed for a better view).}
  \label{fig:baikal_gvd}
\end{figure}

As for the other neutrino telescopes, Baikal-GVD is essentially a 3D array of optical sensors arranged on vertical strings (see Fig.~\ref{fig:baikal_gvd}). 
The Baikal-GVD strings are grouped in clusters, 8 strings per cluster, with 60 m horizontal spacing between the strings.
Each string is 525 m long and holds 36 optical modules (OMs) spaced every 15 m along the string.
Strings are anchored to the lakebed and kept vertical by buoys at their tops, as well as by the buoyancy of the optical and electronics modules.
The OM includes a 10-inch high-quantum-efficiency PMT (Hamamatsu R7081-100), a high voltage unit and front-end electronics, all enclosed in a glass sphere.
The OM is also equipped with calibration LEDs, a tiltmeter/accelerometer and a compass.
The string also holds electronic modules housed in glass spheres, hydrophones for acoustic positioning and LED beacons for calibration \cite{Baikal_calibration,Baikal_positioning}.
All modules are connected with electrical cables for power and data transmission.
All strings of a given cluster are connected to the cluster center modules attached above the central string of the cluster.
The cluster center is connected to the shore station via a dedicated electro-optical cable.
It is responsible for distributing the time synchronization signals to the individual strings, cluster-level triggering and data transmission to shore.
Each cluster acts as an independent detector.
The effective volume of a stand-alone GVD cluster for cascade-like neutrino events with energy above 100 TeV is $\sim$0.05 km$^3$.
A common synchronisation clock allows for the subsequent merging of the physics event data collected from the different clusters.
Additional dedicated strings equipped with high-power pulsed lasers are installed in-between the GVD clusters. These are used for detector calibration and light propagation studies.

The first cluster of Baikal-GVD was deployed in 2016.
Two more clusters were added in 2017 and 2018.
Since 2019 two new clusters are deployed each year.
Today Baikal-GVD consists of 7 clusters, occupying a water volume of $\approx$ 0.35 km$^3$.
As it stands, Baikal-GVD is currently the largest neutrino telescope in the Northern Hemisphere.
The Baikal-GVD construction plan for the period from 2021 to 2024 anticipates the addition of two new clusters every year.

All Baikal-GVD clusters generally show stable operation.
Occasional failures of individual optical or electronics modules (e.g. due to water leaks) are fixed during winter campaigns.
(Each detector string can be recovered and re-deployed without the need to recover the whole cluster.)

\begin{figure}
  \centering
  \includegraphics[height=7.5cm]{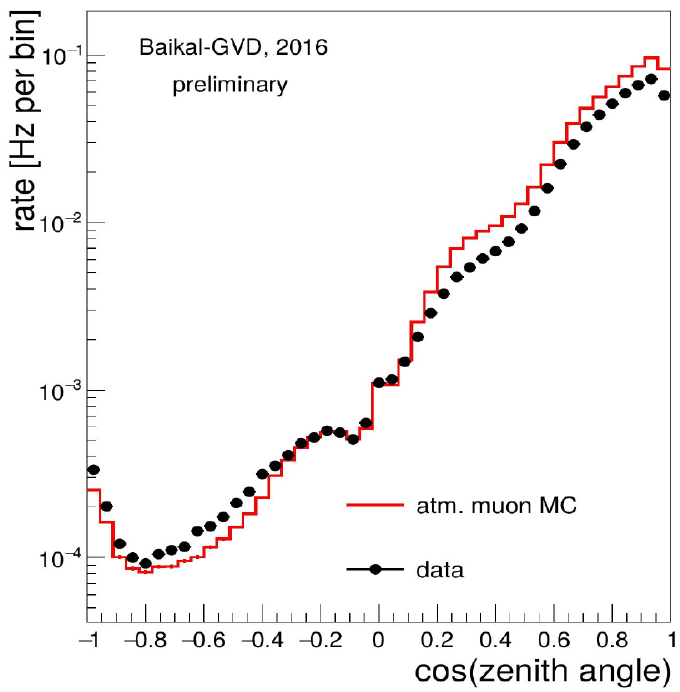}
  \caption{Zenith angle distribution of reconstructed tracks before quality cuts. Baikal-GVD 2016 data.  }
  \label{fig:track_analysis_results_1}
\end{figure}

\begin{figure}
  \centering
  \includegraphics[trim=0 10 20 37,clip,height=7.5cm]{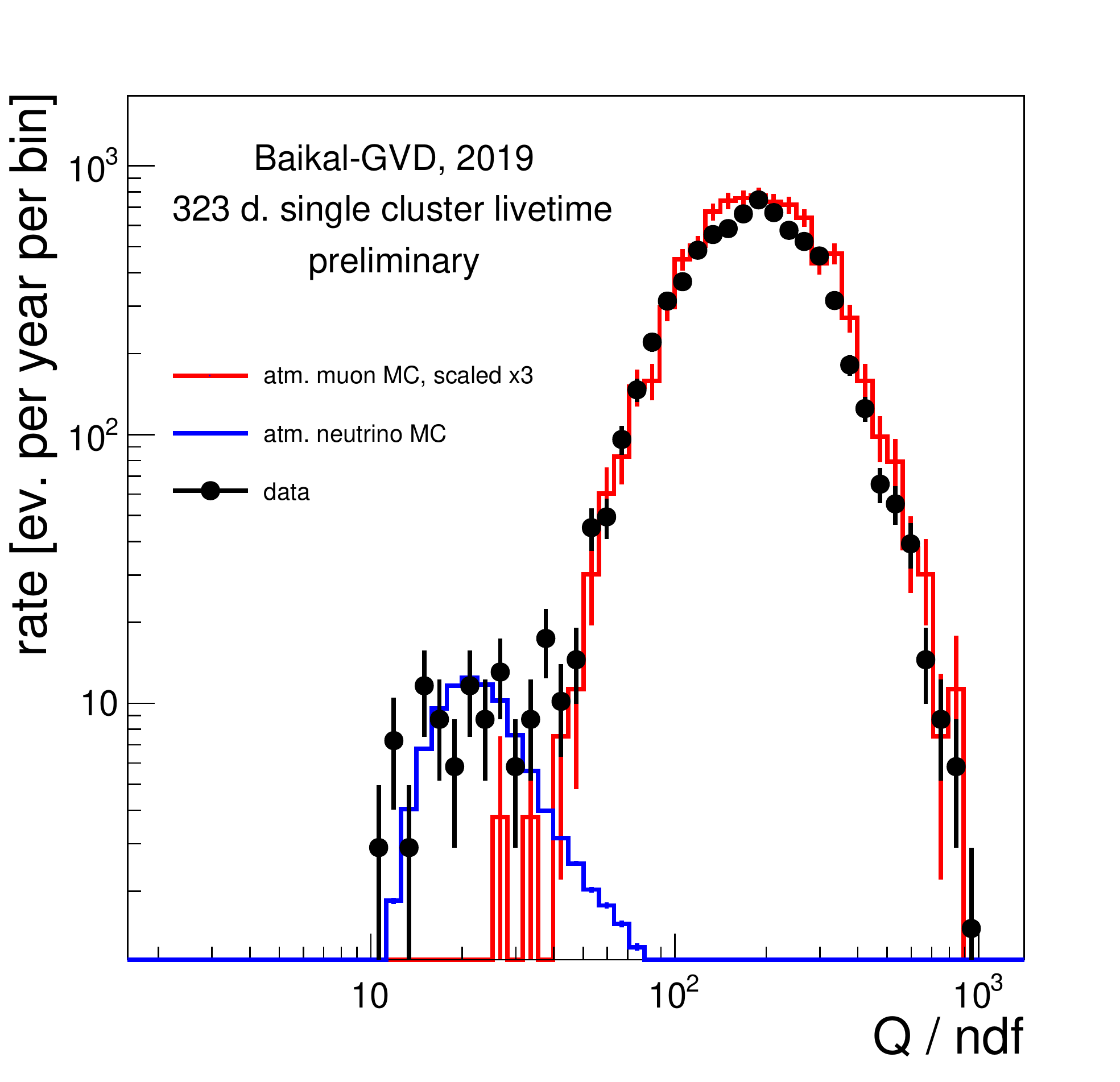}
  \includegraphics[height=7.5cm]{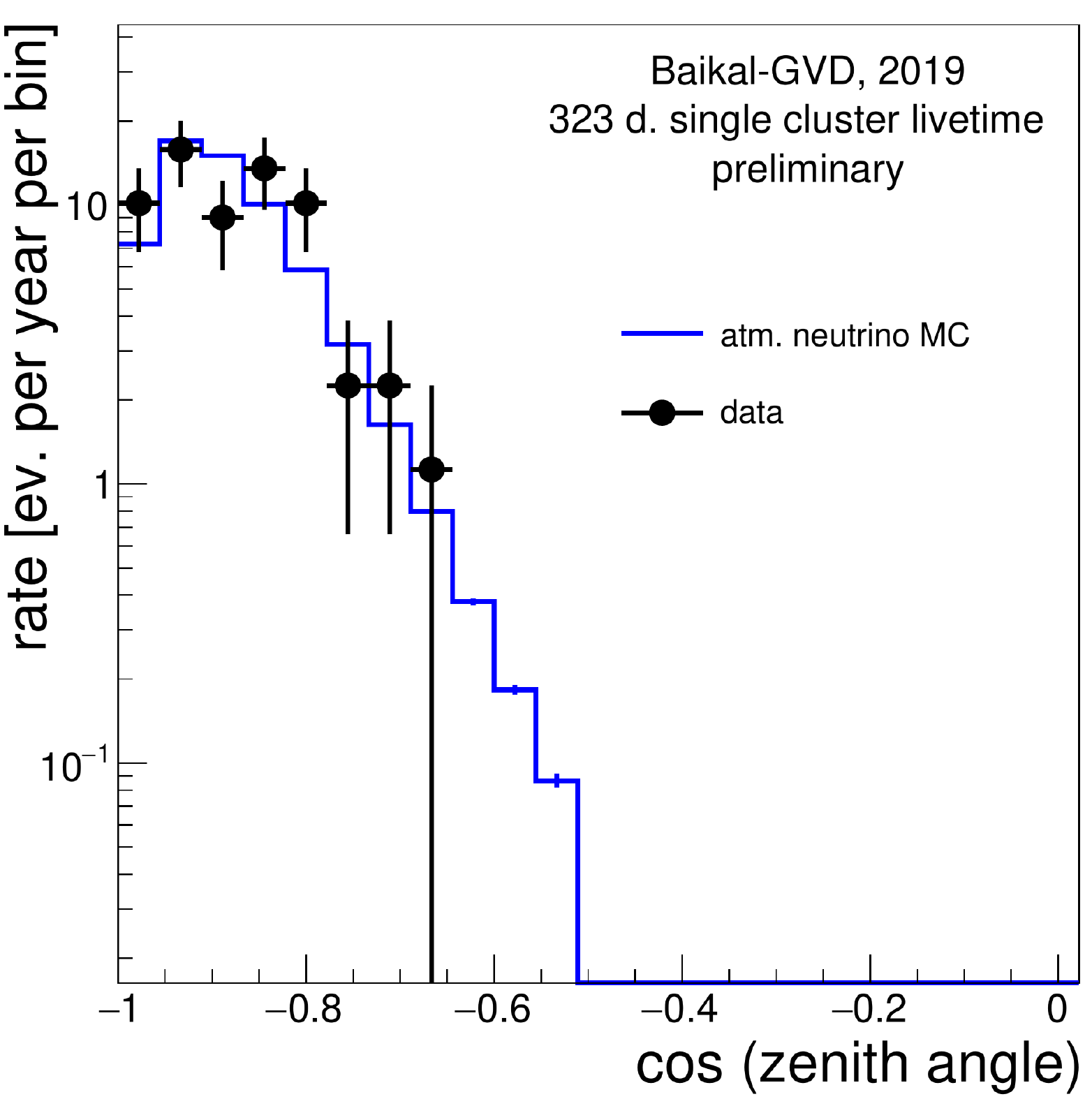}
  \caption{Results from a track analysis of the 2019 Baikal-GVD data.
           Left: the distribution of the fit quality parameter Q/n.d.f. for tracks reconstructed as upward-going. 
           Right: zenith angle distribution of the atmospheric neutrino candidate events after a cut on the fit quality parameter. The expected number of background atmospheric muon events for this sample is $\leq$1 and is therefore not shown.
 }
  \label{fig:track_analysis_results_2}
\end{figure}

Several physics analyses are currently ongoing in Baikal-GVD.
This includes the measurements of atmospheric muons, atmospheric neutrinos and searches for astrophysical neutrino events.
First results from a track analysis of the Baikal-GVD data are shown in Figs.~\ref{fig:track_analysis_results_1} and \ref{fig:track_analysis_results_2}.
These are obtained with a $\chi^2$-like track fit.
The Monte Carlo (MC) expectations for atmospheric muons and atmospheric neutrinos are shown in red and blue, respectively.
The experimental data is shown by black points with statistical error bars.
The atmospheric muon observations (Fig.~\ref{fig:track_analysis_results_1}) are found to be in fair agreement with MC expectations.
The fit quality distribution (Fig.~\ref{fig:track_analysis_results_2}, left) solidly demonstrates the isolation of the upward-going neutrino events from the background of misreconstructed atmospheric muons.
The zenith angle distribution of the atmospheric neutrino candidate events (Fig.~\ref{fig:track_analysis_results_2}, right) is in good agreement with the MC prediction.
The observed deviations from MC expectations in the distributions of atmospheric muon events are under investigation.

\begin{figure}
  \centering
  \includegraphics[height=5.5cm]{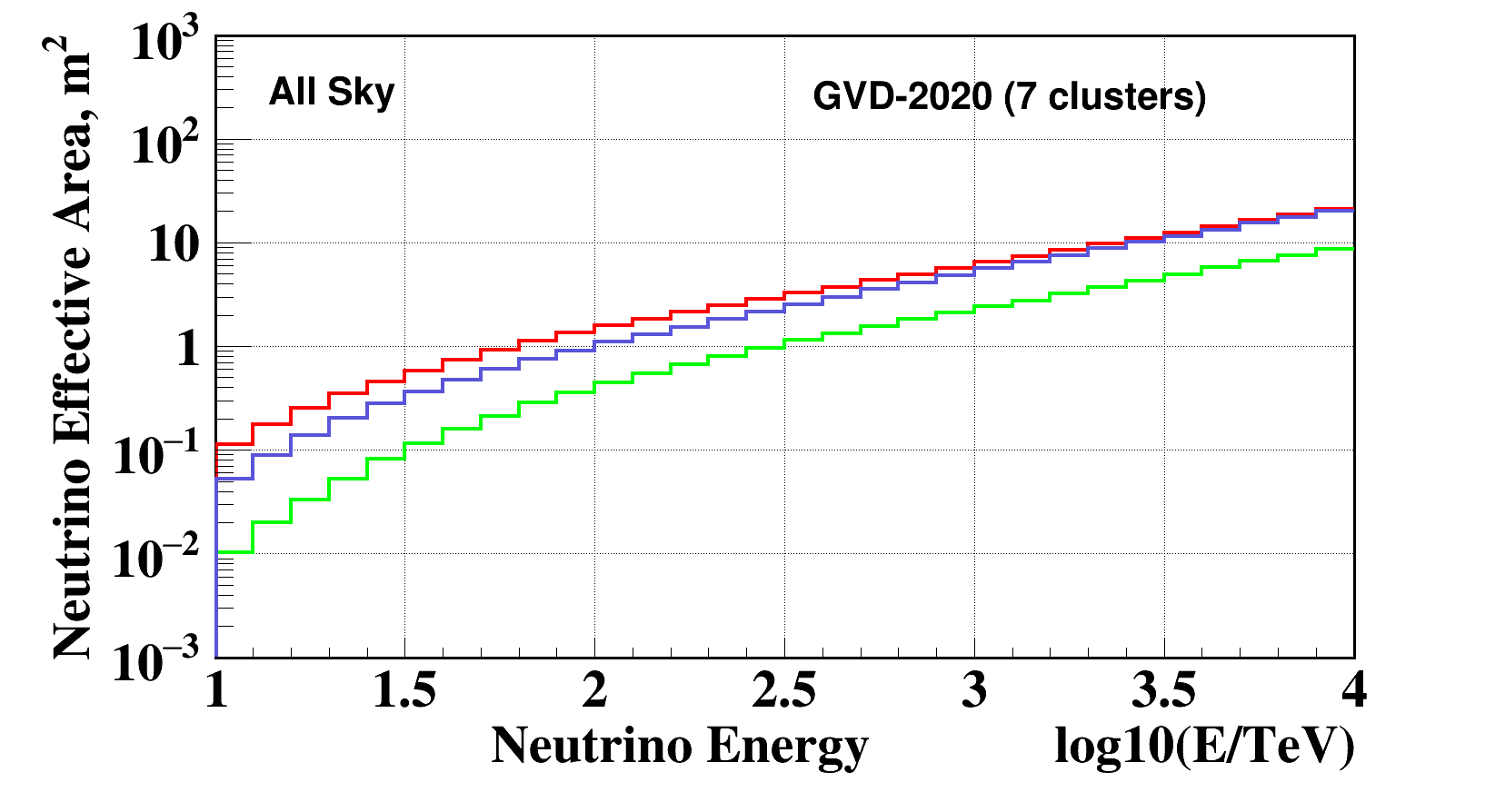}
  \caption{Effective area of the 7-cluster detector (GVD-2020) for high-energy neutrino detection using the cascade analysis (see text).
The red, green, and blue curves (from top to bottom) correspond to $\nu_e$, $\nu_\tau$, and $\nu_\mu$, respectively.
 The Glashow resonance is not included.}
  \label{fig:track_cascade_effarea}
\end{figure}

\begin{figure}
  \centering
  \includegraphics[height=5.5cm]{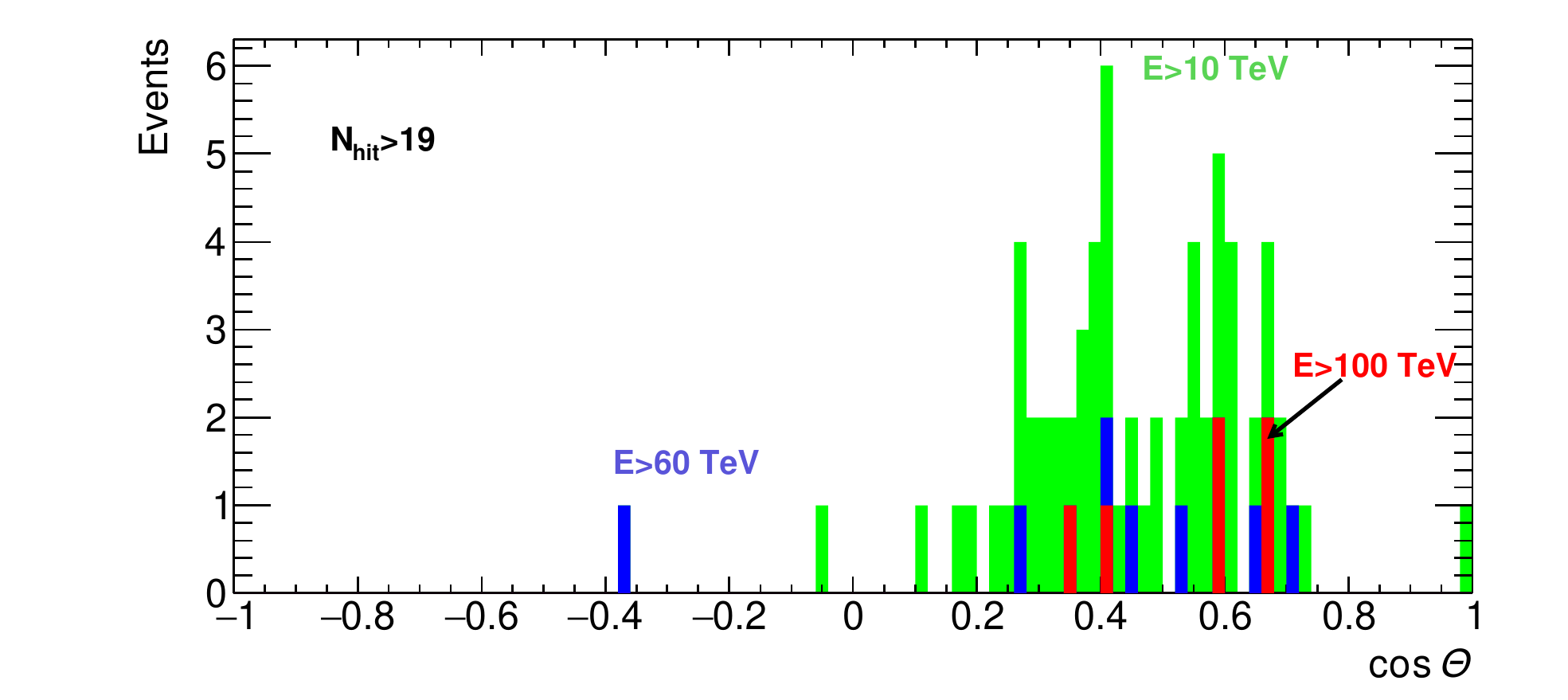}
  \caption{Reconstructed zenith angle distribution of high-energy cascade-like events for a data sample collected with the 7-cluster detector (see text).}
  \label{fig:track_cascade_analysis}
\end{figure}

A high-energy cascade analysis is performed using a likelihood-based fit \cite{Baikal_cascades}.
The projected neutrino effective area for the cascade analysis is shown in Fig.~\ref{fig:track_cascade_effarea}.
The analysis is primarily sensitive to neutrinos and anti-neutrinos of electron and tau flavours.
The reconstructed zenith angle distribution of the observed events is shown in Fig.~\ref{fig:track_cascade_analysis}.
The green, blue and red histograms correspond to events reconstructed with energy above 10 TeV, 60 TeV, and 100 TeV, respectively. With the 10 TeV threshold, the sky above horizon ($cos(\theta)>0$) is expected to be dominated by the background of misreconstructed atmospheric muon events, which is still under evaluation.
For $cos(\theta)<0$ and $E>60$~TeV the observed events are expected to be dominated by isolated cascades from neutrino events,
with the majority of them coming from the astrophysical diffuse neutrino flux.
In this analysis, one such event is observed, compatible with the expectation for the astrophysical neutrino flux (4.2 events per year for the 7-cluster detector).
The event has an estimated energy of $\approx$ 91 TeV.
Upper limits have been derived for some selected sources and multi-messenger alerts, 
e.g. for the gravitational signal GW~170817 \cite{Baikal_multimess}.

\section{Conclusion}
Neutrino astronomy provides a new viewpoint at the non-thermal Universe.
Neutrinos can tell us something about violent processes happening around compact objects, including locations covered from our view by dense matter or photon fields which block electromagnetic radiation.
The discovery of the diffuse neutrino flux by IceCube led to new questions about the origin of the flux.
The discovery of neutrino emission from a first point-like source, TXS 0506+056, is a great success owing to decades of dedicated efforts on detector development and construction.
IceCube observations need to be complemented with observations by similar-size telescopes in the Northern Hemisphere, such as KM3NeT and Baikal-GVD.

The construction of Baikal-GVD has been steadily progressing in the past few years. 
With a 0.35 km$^3$ volume for high-energy cascade detection, Baikal-GVD is currently the largest neutrino telescope in the Northern Hemisphere.
The telescope has been collecting data in partial configurations since 2016.
The Baikal-GVD observations of atmospheric muons are in fair agreement with expectations.
The observed flux of atmospheric neutrino events, using a single-cluster analysis, is in excellent agreement with MC predictions.
First astrophysical neutrino candidates events have been observed using a cascade analysis.

\section*{Acknowledgements}
This work is supported by the Ministry of Science and Higher Education of Russian Federation under the contract 075-15-2020-778 in the framework of the Large scientific projects program within the national project ``Science'' and under the contract FZZE-2020-0017.
We acknowledge the support from RFBR grant 20-02-00400 and grant 19-29-11029,
as well as support by the JINR young scientist and specialist grant 20-202-09.
We also acknowledge the technical support of JINR staff for the computing facilities (JINR cloud).

\end{document}